%% file: Gitor.tex
\documentclass[sigconf, screen]{acmart}

\usepackage{cleveref}
\usepackage{listings}
\usepackage{graphicx}
\usepackage{textcomp}
\usepackage{array}
\usepackage{subfigure}
\usepackage{caption}
\usepackage{multirow}
\usepackage{titlesec}
\usepackage{setspace}
\usepackage{algorithm}
\usepackage{algorithmic}
\usepackage{appendix}
\usepackage{listings}
\usepackage{xcolor}
\usepackage{amsmath}
\usepackage{tabularx}
\usepackage{tikz}
\usepackage{etoolbox}
\usepackage{stfloats}
\usepackage{color}
\usepackage{booktabs}
\usepackage{balance}

\usetikzlibrary{matrix,positioning}
\definecolor{codegreen}{rgb}{0,0.6,0}
\definecolor{codegray}{rgb}{0.5,0.5,0.5}
\definecolor{codepurple}{rgb}{0.58,0,0.82}
\definecolor{backcolour}{rgb}{0.95,0.95,0.92}

\lstdefinestyle{mystyle}{
    backgroundcolor=\color{white},
    basicstyle=\scriptsize\ttfamily,
    breakatwhitespace=false,
    breaklines=true,
    captionpos=b,
    commentstyle=\color{green},
    extendedchars=true,
    frame=single,
    keepspaces=true,
    keywordstyle=\color{blue},
    language=Java,
    numbers=left,
    numbersep=5pt,
    numberstyle=\tiny\color{gray},
    rulecolor=\color{black},
    showspaces=false,
    showstringspaces=false,
    showtabs=false,
    stringstyle=\color{red},
    tabsize=2,
    xleftmargin=2em,
    xrightmargin=2em,
    title=\lstname,
}

\crefname{lstlisting}{listing}{listings}
\Crefname{lstlisting}{Listing}{Listings}

\author{Junjie Shan}
\authornote{Equal contribution}
\email{shanjunjie@westlake.edu.cn}
\affiliation{%
  \institution{Westlake University}
  \city{Hangzhou}
  \country{China}
}

\author{Shihan Dou}
\authornotemark[1]
\email{shihandou@foxmail.com}
\affiliation{%
  \institution{Fudan University}
  \city{Shanghai}
  \country{China}
}

\author{Yueming Wu}
\email{wuyueming21@gmail.com}
\authornote{Yueming Wu is the corresponding author}
\affiliation{%
  \institution{Nanyang Technological University}
  \country{Singapore}
}

\author{Hairu Wu}
\email{hrwu20@fudan.edu.cn}
\affiliation{%
  \institution{Fudan University}
  \city{Shanghai}
  \country{China}
}

\author{Yang Liu}
\email{yangliu@ntu.edu.sg}
\affiliation{%
  \institution{Nanyang Technological University}
  \country{Singapore}
}

\renewcommand{\abstractname}{ABSTRACT}
\renewcommand{\refname}{REFERENCES}

\usepackage{url}

\input{outline/mymacros}

\setcopyright{acmlicensed}
\acmPrice{15.00}
\acmDOI{10.1145/3611643.3616371}
\acmYear{2023}
\copyrightyear{2023}
\acmSubmissionID{fse23main-p1813-p}
\acmISBN{979-8-4007-0327-0/23/12}
\acmConference[ESEC/FSE '23]{Proceedings of the 31st ACM Joint European Software Engineering Conference and Symposium on the Foundations of Software Engineering}{December 3--9, 2023}{San Francisco, CA, USA}
\acmBooktitle{Proceedings of the 31st ACM Joint European Software Engineering Conference and Symposium on the Foundations of Software Engineering (ESEC/FSE '23), December 3--9, 2023, San Francisco, CA, USA}
\received{2023-02-02}
\received[accepted]{2023-07-27}

\begin{document}

\title{Gitor: Scalable Code Clone Detection by Building Global Sample Graph}

\input{outline/abstract}


\maketitle


\input{outline/intro}
\input{outline/backgroud}

\input{outline/system}

\input{outline/evaluation}
\input{outline/discussion}
\input{outline/relatedwork}
\input{outline/conclusion}
\input{outline/acknow}

\balance
\bibliographystyle{ACM-Reference-Format}
\bibliography{output}
\end{document}

%% file: outline/mymacros.tex
\newcommand{\ie}{\textit{i.e.,} }
\newcommand{\eg}{\textit{e.g.,} }

\newcommand\opt[1]{}
\newcommand\find[1]{}

\newcommand{\ls}[1]
   {\dimen0=\fontdimen6\the\font 
    \lineskip=#1\dimen0
    \advance\lineskip.5\fontdimen5\the\font
    \advance\lineskip-\dimen0
    \lineskiplimit=.9\lineskip
    \baselineskip=\lineskip
    \advance\baselineskip\dimen0
    \normallineskip\lineskip
    \normallineskiplimit\lineskiplimit
    \normalbaselineskip\baselineskip
    \ignorespaces
   }

\newenvironment{smalldescription}{
   \setlength{\topsep}{0pt}
   \setlength{\partopsep}{0pt}
   \setlength{\parskip}{0pt}
   \begin{description}
   \setlength{\leftmargin}{.2in}
   \setlength{\parsep}{0pt}
   \setlength{\parskip}{0pt}
   \setlength{\itemsep}{0pt}}{\end{description}}



%% file: outline/abstract.tex
\begin{abstract}
Code clone detection is about finding out similar code fragments, which has drawn much attention in software engineering since it is important for software maintenance and evolution.
Researchers have proposed many techniques and tools for source code clone detection, but current detection methods concentrate on analyzing or processing code samples individually without exploring the underlying connections among code samples.

In this paper, we propose \emph{Gitor} to capture the underlying connections among different code samples.
Specifically, given a source code database, we first tokenize all code samples to extract the pre-defined \emph{individual information} (\eg \emph{keywords}). 
After obtaining all samples' individual information, we leverage them to build a large \emph{global sample graph} where each node is a code sample or a type of individual information.
Then we apply a node embedding technique on the global sample graph to extract all the samples' vector representations. After collecting all code samples' vectors, we can simply compare the similarity between any two samples to detect possible clone pairs. 
More importantly, since the obtained vector of a sample is from a global sample graph, we can combine it with its own code features to improve the code clone detection performance.
To demonstrate the effectiveness of \emph{Gitor}, we evaluate it on a widely used dataset namely BigCloneBench.
Our experimental results show that \emph{Gitor} has higher accuracy in terms of code clone detection and excellent execution time for inputs of various sizes (1–100 MLOC) compared to existing state-of-the-art tools.
Moreover, we also evaluate the combination of \emph{Gitor} with other traditional vector-based clone detection methods, the results show that the use of \emph{Gitor} enables them detect more code clones with higher F1.

\end{abstract}

%



%% file: outline/intro.tex
\section{INTRODUCTION}
Code clone, also known as duplicate code or similar code, refers to the existence of two or more identical or similar source code fragments. 
Numerous empirical studies~\cite{10.5555/832303.836911, Kim2005AnES, roy2008nicad} have shown that code cloning widely exists in different open source or closed source code bases. 
For example, \cite{10.5555/832303.836911,777743} detected 22.3\% of code clones in Linux system, Kamiya et al. found 29\% code clones in JDK, and even up to 50\% code clones in some software systems~\cite{walenstein2007software}. 
Widespread code cloning has helped the development of software systems to a certain extent and can have positive benefits \cite{supporting,rattan2013software}.
However, many studies have pointed out that a large number of code clones can have a negative impact on software systems maintenance \cite{4658071,5070547,thummalapenta2010empirical}, since it may introduce bugs or vulnerabilities. 
Therefore, the automatic detection of code clones has attracted wide attention in the field of software engineering.

According to the syntactic or semantic similarity of code clones, Bellon et al. classified code clones into four types \cite{roy2007type1_4, bellon2007type1_4}: textual similarity (type 1), lexical similarity (type 2), syntactic similarity (type 3), and semantic similarity (type 4). 
From type 1 to type 4, the similarity of cloned codes gradually decreases and the difficulty of detection gradually increases. 
A number of code clone detection method has been proposed~\cite{svajlenko2014big,kamiya2002ccfinder,sajnani2016sourcerercc,white2016rtvnn,zhang2019astnn,jiang2007deckard,wang2017ccsharp}. 
For example, a state-of-the-art token-based method namely \emph{SourcererCC} \cite{sajnani2016sourcerercc} is designed to capture the tokens’ overlap similarity among different methods to detect Type-1 to Type-3 clones.
In practice, token-based techniques are unable to handle Type-4 clones (\ie semantic clones) due to a lack of respect for program semantics. 
To mitigate the issue, researchers use program analysis to distill the semantics of code fragments into tree or graph representations (\eg abstract syntax tree and control flow graph) and apply tree or graph matching to quantify the similarity between different codes. 
Empirical studies \cite{krinke2001duplix, komondoor2001pdgdup, wang2017ccsharp} have shown that tree-based and graph-based code clone detectors can achieve better performance on semantic code clone analysis.
However, due to the complexity of tree and graph structures, they are unable to scale to large programs.
Given that large-scale clone detection is essential to daily software engineering activities such as code search \cite{keivanloo2011code}, mining library candidates \cite{ishihara2012library}, and license violation detection \cite{german2009license1, koschke2012license2}, there is an increasing need for a scalable technique to detect code clones.

In this paper, we propose a novel code clone detection method leveraging global graph built across code samples. We find that almost all current code clone detection methods focus on extracting the features from source code directly while ignoring the potential underlying connections among different code samples.
To achieve scalable and accurate code clone detection, we consider extracting these connections to build ``bridges'' between code samples (\ie global graph) and using them to detect code clones.
Specifically, we mainly address two challenges in our paper.
\begin{itemize}
\item{\emph{How to build the global graph from source code and represent it properly to retain code details?}}
\item{\emph{How to utilize the global graph across different source code samples to efficiently and accurately detect code clones?}}
\end{itemize}

To tackle the first challenge, we choose keyword tokens along with side information as the individual information to represent the source code samples.
In detail, since the programming language of our experimental dataset is Java, we leverage the reserved words of Java as keyword tokens. 
Meanwhile, to better capture the code details, we also extract another kind of information (\ie side information) such as the maximum depth of brackets and the number of loops.
Because the extraction of keywords and side information can be achieved by simple lexical analysis, we can complete scalable code clone analysis.

To address the second challenge, we use keywords and side information as the ``bridge'' to connect different code samples.
Specifically, we build a global sample graph to represent the underlying connections between all samples.
Each node in the graph represents a code sample or a kind of individual information (\ie keywords or side information).
Each edge indicates whether a code sample contains such individual information.
After constructing the global graph, we perform a node embedding technique on it to convert all code samples into corresponding vector representations.
Given generated vectors, we can calculate the \emph{cosine similarity} of two samples and quickly identify whether they are clone pairs.


We implement a prototype system, \emph{Gitor}, and evaluate it on a widely used dataset, namely BigCloneBench \cite{big, svajlenko2014big}. 
Our evaluation results show that \emph{Gitor} is superior to six state-of-the-art comparative systems including \emph{SourcererCC} \cite{sajnani2016sourcerercc}, \emph{CCFinder} \cite{kamiya2002ccfinder}, \emph{Nicad} \cite{roy2008nicad}, \emph{Deckard} \cite{jiang2007deckard}, \emph{CCAligner} \cite{wang2018ccaligner}, \emph{Oreo} \cite{saini2018oreo}, \emph{LVMapper} \cite{wu2020lvmapper}, and \emph{NIL} \cite{nakagawa2021nil}. 
Moreover, we can also combine the code sample representation vector generated by \emph{Gitor} with feature vector obtained from source code directly by three traditional vector-based tools (\ie word2vec \cite{word2vec}, doc2vec \cite{doc2vec}, and code2vec \cite{alon2019code2vec}), the results indicate that the combination make them detect more clones with higher F1.
Finally, we examine the scalability of \emph{Gitor} on various sizes of code.
Evaluation results report that \emph{Gitor} has the ability to analyze 100 million lines of code, with the shortest execution time compared to \emph{SourcererCC}, \emph{CCFinder}, \emph{Nicad} , \emph{Deckard}, \emph{CCAligner}, \emph{Oreo}, \emph{LVMapper}, and \emph{NIL}.

\par In summary, this paper makes the following contributions:
\begin{itemize}
  \item We propose a novel method to detect code clones by building a global sample graph using keywords and side information.
  The constructed global graph can capture the underlying connections between different source code samples.
  \item We design a prototype system namely \emph{Gitor} and conduct evaluations on a widely used dataset (\ie BigCloneBench \cite{big}).
  Experimental results suggest that \emph{Gitor} outperforms 
  \emph{SourcererCC}, \emph{CCFinder}, \emph{Nicad} , \emph{Deckard}, \emph{CCAligner}, \emph{Oreo}, \emph{LVMapper}, and \emph{NIL}
  and \emph{Gitor} is adept at handling the challenges posed by the big scale of code.
\end{itemize}

\par \noindent \textbf{Paper organization.} The remainder of the paper is organized as follows.
Section 2 explains the background and motivation.
Section 3 introduces our system. 
Section 4 reports the experimental results. 
Section 5 discusses the future work. 
Section 6 describes the related work.
Section 7 concludes the present paper.



%% file: outline/backgroud.tex
\section{DEFINITION AND MOTIVATION}
\subsection{Definitions}
The paper utilizes the well-accepted definitions of code clones and clone types as follows:

\begin{lstlisting}[caption=Original (Func \#0), label=lst:original]
public static int fib(int i){
    int f1=0, f2=1, c=0;
    if((i == 0) || (i == 1)) return i;
    for (int j =2; j<=i; j++){
        c=f1+f2; f1=f2; f2=c;
    }
    return c;
}
\end{lstlisting}

\begin{lstlisting}[caption=Type-1 (Func \#1), label=lst:type1]
public static int fib(int i){
    int f1=0, f2=1, c=0;
    if((i == 0) || (i == 1)) return i;
    for (int j =2; j<=i; j++){
        c=f1+f2; f1=f2; f2=c;
    }
    return c;
}
\end{lstlisting}

\begin{lstlisting}[caption=Type-2 (Func \#2), label=lst:type2]
public static int fib(int num){
    int f1=0, f2=1, c=0;
    if((num == 0) || (num == 1)) return num;
    for (int j =2; j<=num; j++){
        c=f1+f2; f1=f2; f2=c;
    }
    return c;
}
\end{lstlisting}


\begin{lstlisting}[caption=Type-3 (Func \#3), label=lst:type3]
public static int calFib(int num){
    int fib1=0, fib2=1;
    int t=0;
    if((num == 1) || (num == 0)) return num;
    for (int k =2; k<=num; k++){
        t=fib1+fib2; fib1=fib2; fib2=t;
    }
    return t;
}
\end{lstlisting}

\begin{lstlisting}[caption=Type-4 (Func \#4), label=lst:type4]
public static long calFib(long number){
    long f1=0, f2=1, c=0;
    switch(number){
        case 0:
            return 0;
        case 1:
            return 1;
        default:
            break;
    }
    while(number>=2){
        c=f1+f2; f1=f2; f2=c;
        number--;
    }
    return c;
}
\end{lstlisting}




In our paper, we use the following widely used definitions \cite{roy2007type1_4, bellon2007type1_4} of code clone types.

\begin{itemize}
  \item \textbf{Type-1 (textual similarity)}: Identical code fragments, except for minor differences in white-space, layout, or comments.
  \item \textbf{Type-2 (lexical similarity)}: Structurally identical code fragments, in addition to Type-1 clone differences, there might be some differences in identifier names and literal values.
  \item \textbf{Type-3 (syntactic similarity)}: Modified similar code fragments that differ at the statement level. Besides the Type-1 and Type-2 clone, the fragments might have statements added, modified and/or removed compared to each other.
  \item \textbf{Type-4 (semantic similarity)}: Dissimilar code fragments with the same functionality but implemented in a syntactically different way.
\end{itemize}

\par To elaborate on different types of clones, \cref{lst:original,lst:type1,lst:type2,lst:type3,lst:type4} present examples from Type-1 to Type-4 clones. 
The original code is used to compute the Fibonacci number given the order. 
The Type-1 clone (starting in line \#11) is identical to the original code. 
The Type-2 clone (starting in line \#21) differs only in identifiers name (\ie $m$ and $n$ instead of $a$ and $b$). Obviously, the two types mentioned above are easy to detect. 
The Type-3 clone (starting in line \#31) is syntactically similar but differs at the statement level. 
The first line in Type-3 (line \#42) is totally different from the origin code. The method name and types of parameters are all changed. 
In addition, it calculates the greatest common divisor in a similar but not identical way. Detecting Type-3 clones is harder than the previous two types. 
Finally, the Type-4 clone (starting in line \#42) iterates to compute the greatest common divisor which is a completely different way. 
Its lexical and syntactic are dissimilar to the original method. 
Therefore, it requires an in-depth understanding of code fragments to detect Type-4 clones.

\subsection{Motivation}
To illustrate the key insight of our proposed method, we leverage Fun \#0 and its corresponding type 3 and type 4 clones (\ie Fun \#3 and Fun \#4) as our analysis targets. 
As shown in Listing 1, those examples are all used to calculate the Fibonacci number of the given order. 
According to the definition of code clone, the clone pair \emph{Fib$_0$.java} and \emph{Fib$_3$.java} are classified as Type-3 clone (\ie syntactic similarity) since they differ at the statement level.
The clone pair \emph{Fib$_0$.java} and \emph{Fib$_4$.java} are classified as a Type-4 clone (\ie semantic clone) because they have syntactically dissimilar code to implement the same functionality.

\subsubsection{SourcererCC.}
\par We start with illustrating how the widely used clone detection tool \emph{SourcererCC} \cite{sajnani2016sourcerercc} (\ie one of the state-of-the-art token-based clone detectors) detects possible clone pairs by calculating the similarity of each pair.
\emph{SourcererCC} \cite{sajnani2016sourcerercc} utilizes the \emph{Overlap} of two source code blocks to compute the similarity since it intuitively captures the notion of overlap among different code blocks.
For example, given two code blocks \emph{C$_1$} and \emph{C$_2$}, the overlap similarity \emph{S}(\emph{C$_1$}, \emph{C$_2$}) is calculated as the number of tokens shared by \emph{C$_1$} and \emph{C$_2$}.

\begin{center}
$\emph{S}(\emph{C$_1$}, \emph{C$_2$}) = \left | \emph{C$_1$}\bigcap \emph{C$_2$} \right |$
\end{center}

Given the threshold $\theta$ and the maximum number of tokens $T = max(|C_1|,|C_2|)$, a pair of code blocks is considered as a clone pair when the ratio of overlap similarity and $T$ is greater than the threshold $\theta$.
\begin{center}
    $\frac{S \left( C_1, C_2\right)}{T} \geq \theta$
\end{center}

\subsubsection{Keywords.}
To achieve a more accurate clone detection, we need to extract reliable information to represent the source code, preferably some kind of global information that can reflect the connections between source code samples rather than analyze the information from code samples individually. So, we will extract the individual information from each code sample by extracting keywords and build a global sample graph that connects all code samples.
\begin{figure}[htbp]
\centerline{\includegraphics[width=0.4\textwidth]{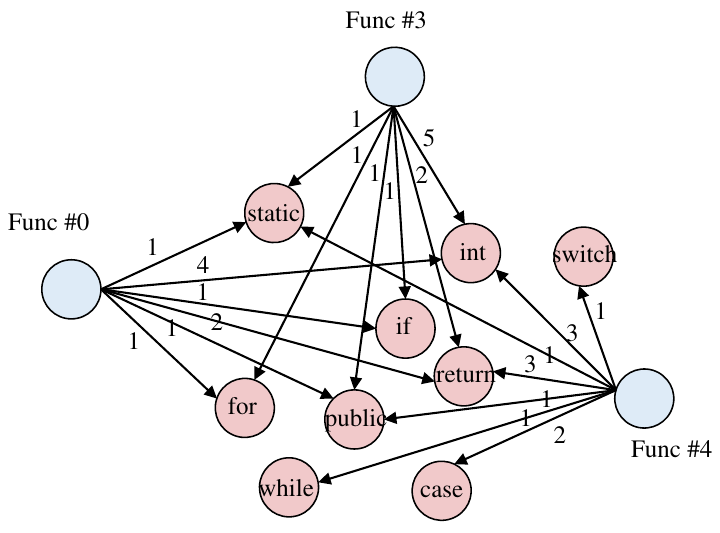}}
\caption{A global graph of Func\#0, Func\#3, and Func\#4.}
\label{fig:graph}
\vspace{-1em}
\end{figure}

We first tokenize the source code to get the sequences of tokens of Func\#0, Func\#3, and Func\#4. 
Then we choose only the reserved words in Java as the keywords instead of all tokens to represent code samples since the reserved words are used in all Java source code samples. 
After extracting keywords from the above code samples, we construct a weighted directed graph with the frequency of keywords as weight as illustrated in Figure \ref{fig:graph}. 
Each blue node represents a code sample, each red node represents a keyword, and the weight from blue nodes to red nodes is the frequency of keyword in the corresponding code sample.

\subsubsection{Node Embedding.}
In order to obtain the similarity of Fun\#0, Fun\#3, and Fun\#4 in Figure \ref{fig:graph}, we first use node embedding methods to convert them into their vector representations. These node embedding algorithms typically aim to capture the structural information and relationships between nodes and covert the graph structure and node attributes into representation vectors, which can preserve the underlying similarity among nodes \cite{grover2016node2vec, palumbo2018knowledge}. In this paper, we mainly consider two different embedding methods, namely node2vec~\cite{grover2016node2vec} and ProNE~\cite{zhang2019prone}, since they support the embedding of weighted graph. However, to achieve the scalability, we choose ProNE as our embedding method because it is faster, more scalable, and more effective than node2vec~\cite{LIU202143}.
So, we use ProNE~\cite{zhang2019prone} to map the code samples into vectors, which can be used to calculate the similarity among different functions.

\subsubsection{Similarity Evaluation.}
\par We calculate the similarity of above two code blocks using the method mentioned in \emph{SourcererCC}.
It shows that the number of tokens in Func\#0 and Func\#3 is 73 and 74, respectively. Then the same tokens shared by Fun\#0 and Func\#3 are obtained for computing the overlap similarity. We observe that there are 18 same tokens shared by these two code blocks, which means the overlap similarity of Func\#0 and Func\#3 is 18/74=0.24. If similarity threshold in \emph{SourcererCC} is set to 70\%, which means that \emph{SourcererCC} reports two methods as a clone pair only when the ratio of number of shared tokens and maximum number of tokens of them is larger than 70\%. In this case, \emph{SourcererCC} will cause a false negative by reporting Func\#0 and Func\#3 as a none clone pair. Also, the similarity between Func\#0 and Func\#4 according to \emph{SourcererCC} is 0.23. 
Now, we conduct node embedding on the graph shown in Figure~\ref{fig:graph}, and then we get the vector representations of Fun\#0, Fun\#3, and Fun\#4, which are used to calculate the similarity. After node embedding, the similarity between Func\#0 and Func\#3 is 0.99 and the similarity between Func\#0 and Func\#4 is 0.65, suggesting that the similarity among these clone pairs is significantly improved.


Based on the observation, we propose a novel code clone detection framework by considering the global relationships between different functions.

%% file: outline/system.tex
\section{APPROACH}
\begin{figure*}[htp]
\centering
\includegraphics[width=0.9\textwidth]{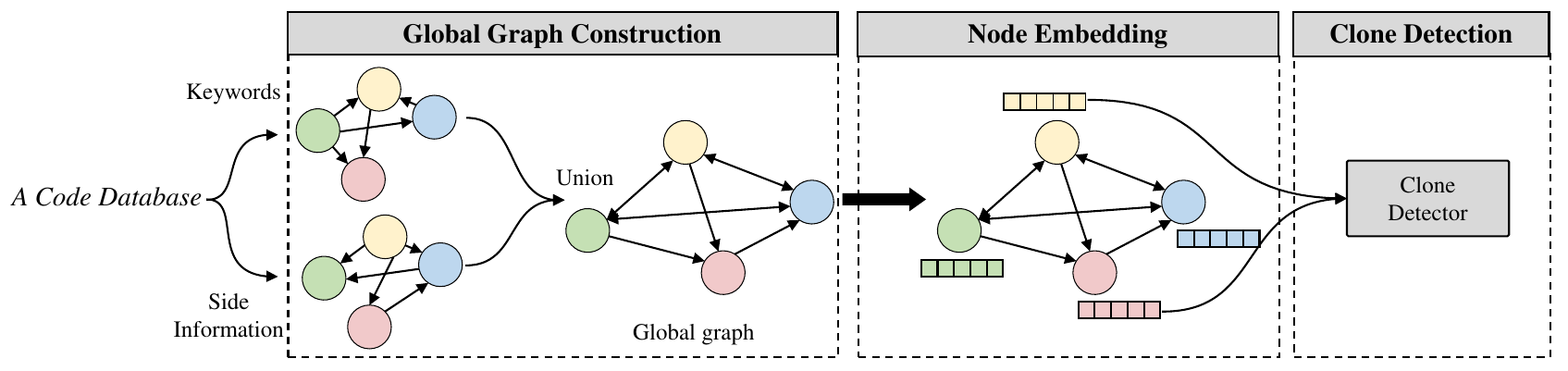}
\caption{System overview of \emph{Gitor}}
\vspace{-1em}
\label{fig:system}
\end{figure*}

\par In this section, we introduce our proposed system, namely \emph{Gitor}. 

\subsection{System Overview}
\par As shown in Figure ~\ref{fig:system}, \emph{Gitor} consists of three main phases: \emph{Global Graph Construction}, \emph{Node Embedding}, and \emph{Clone Detection}.

\begin{itemize}
  
\item \textbf{\emph{Global Graph Construction}}:
We first apply lexical analysis to extract the \emph{individual information} including \emph{keywords} and \emph{side information} of a code sample with corresponding weights.
Then a global sample graph is built by using these information where each node represents a sample or a type of \emph{individual information}.

  
\item \textbf{\emph{Node Embedding}}: Given a graph of the whole code base, we use a node embedding technique on the global graph and output the vectors of each node with chosen dimension. 
The input is a weighted global sample graph, and the outputs are vectors of all samples in the code base.
  
\item \textbf{\emph{Clone Detection}}: After the generation of vectors, we have two ways to detect potential clone pairs. 
First, we can simply calculate the cosine similarity of a pair of samples to identify code clones. 
Second, we can combine \emph{Gitor} with other vector-based clone detection methods, which will boost the performance of clone detection.
\end{itemize}

\subsection{Global Graph Construction}

\subsubsection{Individual Information Extraction.}

\par In this paper, we aim to combine the connection capture capability of graph embedding methods with the scalability of token-based methods. 
Therefore, we first conduct tokenization on the source code to extract the keywords and side information from the source code. 
Since our experiments are done on the BigCloneBench dataset \cite{svajlenko2014big}, we tokenize the \emph{java} source code based on a java parse tool, namely \emph{javalang} \cite{javalang}. 
We choose the \emph{Java} reserved words as \emph{keywords} along with five types of \emph{side-information} as \emph{individual information}. 
For example, take the Func \#0 from Listing~\ref{lst:original}, the keywords and corresponding weights of Func \#0 is \emph{\{public: 1, static: 1, int: 4, if: 1, return: 2, for: 1\}}. 
Moreover, the five different types of \emph{side information} are as follows:
\begin{itemize}
    \item Maximum Nesting Depth of the Curly Brackets (\emph{MNDCB}): The number of maximum depth of nested curly brackets. For example, the \emph{MNDCB} of Func \#0 is 2.
    \item Maximum Number of Parallel Curly Brackets (\emph{MNPCB}): The number of maximum parallel curly brackets with depth 2. For example, the \emph{MNPCB} of Func \#0 is 1.
    \item Loop-Repetition Information (\emph{LRI}): The number of loop functions used in the code, including for-loop and while-loop. For example, the \emph{LRI} of Func \#0 is 1.
    \item Flow-Control Information (\emph{FCI}): The number of flow-control functions used in the code, including if-else and switch-case. For example, the \emph{FCI} of Func \#0 is 1.
    \item Numerical Declaration Information (\emph{NDI}): The number of numeric variables declared in the code, including int, double, float, byte, short and long declaration. For example, the \emph{NDI} of Func \#0 is 4.
\end{itemize}

For \emph{MNDCB}, we utilize a depth counter, adjusting it with every encountered curly bracket—incrementing for each opening and decrementing for each closing, subsequently noting the peak depth. For \emph{MNPCB}, we discern parallel curly brackets at distinct depths by counting sequential opening and closing pairs. Loop-related tokens like for and while contribute to the \emph{LRI} tally. Similarly, flow-control tokens such as if and switch are counted for \emph{FCI}. The \emph{NDI} is ascertained by enumerating numeric variable declaration tokens like int and double. This token-based methodology offers a nuanced perspective on the code's structure and semantics.

In this paper, we chose these types of \emph{side information} because they can provide additional information about the structure and complexity of the code samples and can help to identify clones that might not have been detected by keyword-based methods alone. 
Also, using such information can improve the scalability of code clone since it reflects the code structure without processing and comparing the whole code sample. 
For example, \emph{LRI} represents the number of for-loop declarations and while-loop declarations since they have similar functionality in Java, and the substitution of these two loop functions for each other is often found in clone samples.
Also, the types of \emph{side information} might differ according to the programming language of code samples. 
For instance, the functionality of curly brackets in Python is different from that of curly brackets in Java. 
So, the types of \emph{side information} should be carefully selected according to the different programming languages of code samples. 
Moreover, different combinations of \emph{side information} may affect the detection performance slightly and the main goal of this paper is not about finding the optimal combination of different types of \emph{side information}, so we will use all of these five types in the following paper.

In our study on code clone detection, we discern the importance of both \emph{keywords} and \emph{side information}. Keywords are the reserved words directly extracted as tokens from the code, serving as foundational markers of code content. On the other hand, \emph{side information} delves deeper, encapsulating structural metadata such as bracket depth and loop count. To maximize the potential of both elements, we construct individual graphs for each. These are then amalgamated into a singular, comprehensive global graph, establishing diverse connections between code samples. This method not only merges lexical content with structural nuances but also provides a more robust framework, enhancing the precision in detecting code clones by highlighting intricate relationships and similarities between code snippets.

After extracting \emph{keywords} and \emph{side information}, we can get a sequence of tokens with corresponding weight, which is the frequency in this case.

\begin{figure}
\centering
\subfigure{
\centering
\begin{minipage}[hb]{0.4\textwidth}
\includegraphics[width=0.9\textwidth]{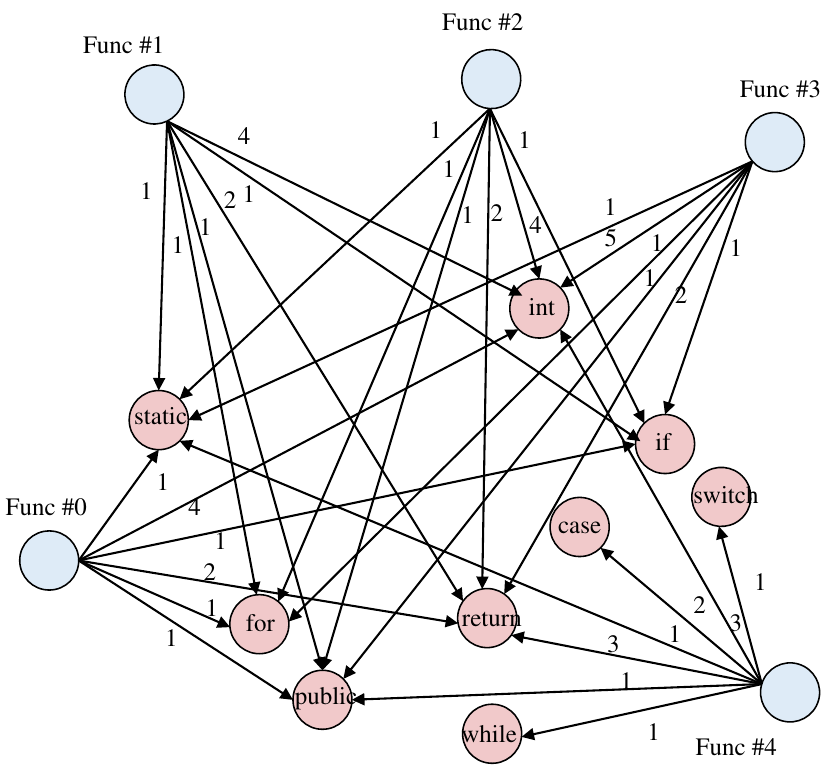}
\end{minipage}
}
\caption{A global graph of Func $\#$0-4 constructed by keywords.}
\label{fig:graph1}
\vspace{-2em}
\end{figure}

\begin{figure}
\centering
\subfigure{
\centering
\begin{minipage}[hb]{0.4\textwidth}
\includegraphics[width=\textwidth]{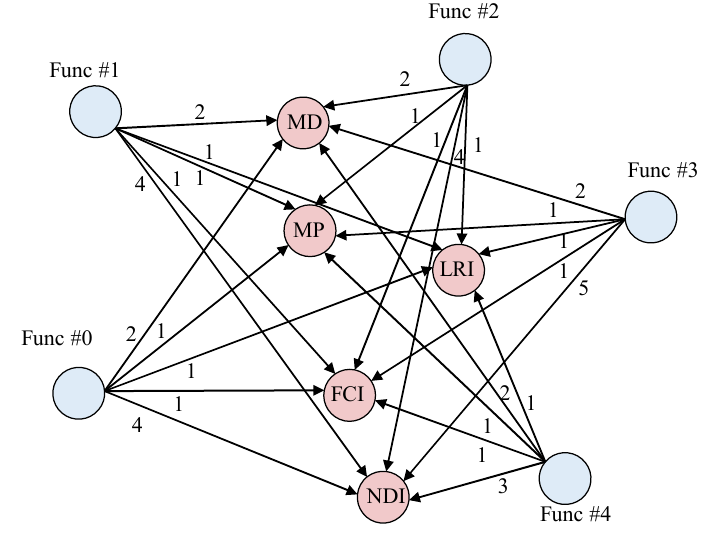}
\end{minipage}
}
\caption{A global graph of Func $\#$0-4 constructed by side information.}
\label{fig:graph2}
\vspace{-2em}
\end{figure}

\subsubsection{Global Graph Construction.}

Nowadays, the graph is an important kind of representation to encode relation structure, which is used in many domains (\ie social networks, citation networks, function call diagrams, etc.). 
The nodes and edges can represent the objects and relationships respectively. 
Evaluation of similarity between two nodes based on the graph structure has a wide range of applications, such as social networks analysis, knn, graph clustering, etc. 
Therefore, instead of simply comparing the similarity using the overlap \emph{keywords} and \emph{side information} of two samples, we first use \emph{keywords} extracted from the last step to build a graph representing the whole code base.
To better illustrate this phase in \emph{Gitor}, we choose the samples in Listing 1 as an example and present a clearer description in Figure \ref{fig:graph1}. 
As illustrated in Figure \ref{fig:graph1}, the blue nodes represent Function $\#$0-4 and the red nodes represent the keywords. 
The weight on the directed edges from functions to keywords is the frequency of keywords appearing in the functions.

To better capture the program details of source code, we also select another kind of information.
Specifically, we construct another graph using \emph{side information} defined above, where the blue nodes represent function as well, but the red nodes represent \emph{side information} types, and the weight is the count of corresponding \emph{side-information}, as illustrated in Figure \ref{fig:graph2}. 
After obtaining two graphs using \emph{keywords} and \emph{side information}, we merge them by merging the nodes that have the same labels to build one larger global sample graph which will be embedded and used to calculate the similarity of any two functions.

\par In short, the input of graph construction is a code database containing many code samples (\ie functions) and the output is a large global sample graph.

\begin{figure*}[htp]
\centering
\includegraphics[width=0.8\textwidth]{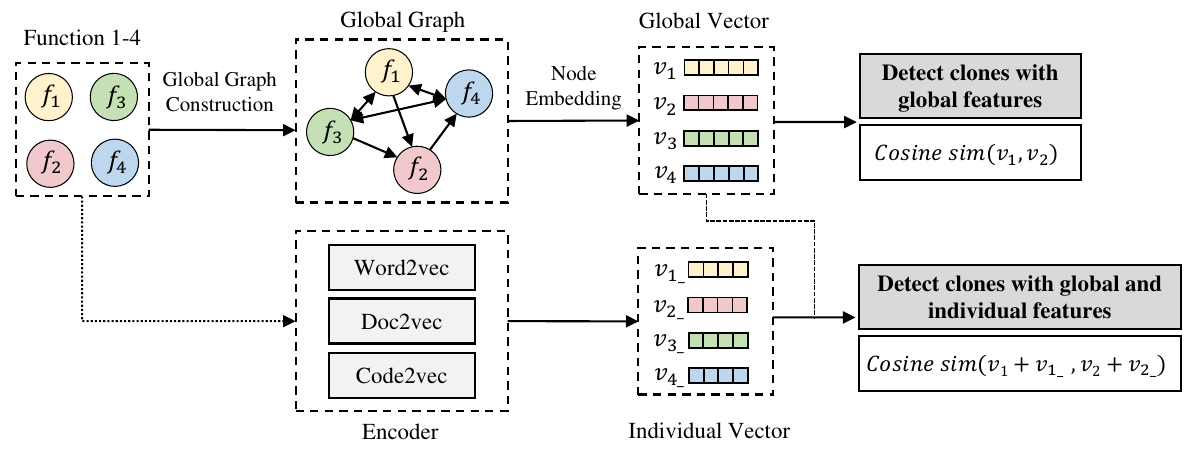}
\caption{Two applications of \emph{Gitor}. The first is to detect code clones using global features and the second is to combine global features with individual features to detect code clones.}
\label{fig:app}
\end{figure*}

\subsection{Node Embedding}

Graph is a commonly used type of information representation in complex systems and can represent many complex relationships in real-life scenarios, such as social networks \cite{newman2002random}, crime networks \cite{huang2018deepcrime}, traffic networks \cite{zheng2020gman}, etc. 
Graph analysis is used to dig deeper into the intrinsic features of graph data, however, since the graph is non-Euclidean data, traditional data analysis methods generally have high computational effort and spatial overhead. 
Graph embedding is an effective method to solve the graph analysis problem, which transforms the original graph data into a low-dimensional space and preserves key information, thus improving node classification, link prediction, and graph analysis. 
It can improve the performance of the tasks like node classification, link prediction, and node clustering by retaining key information from the graph. 
Deep Learning-based methods among different graph embedding methods have demonstrated promising results due to their capability of automatically discovering underlying connections and identifying useful representations from the complex graph structures. 
For instance, deep learning with random walk (\ie DeepWalk~\cite{perozzi2014deepwalk} and Node2vec~\cite{grover2016node2vec}) can leverage the neighborhood structure by sampling paths on the graph automatically.

Graph embedding methods are feature representation learning methods, exploiting the graph structure to transform each node of the graph into a low-dimensional vector while preserving neighborhood similarity, semantic information, and community structure among nodes \cite{dai2019adversarial}. 
The obtained vector representations can be utilized by a wide range of tasks such as link prediction \cite{sun2019atp}, node classification \cite{tang2015line}. 
So, the node embedding method can capture the global connections among nodes in the graph, which means it can capture the underlying similarity property among functions from a holistic perspective than analyzing them individually. 
In this paper, we choose ProNE \cite{zhang2019prone} since it is a fast and effective method that combines the benefits of various embedding methods while remaining time-efficient~\cite{LIU202143}. 
Moreover, we conduct the embedding with different vector sizes (\ie d = 16, 32, 64, 128) on our chosen dataset, BigCloneBench\cite{svajlenko2014big}, to find the optimal embedding dimension for clone detection.

In brief, the input of node embedding is the graph constructed before, and the outputs are vectors of all nodes in the graph with the pre-defined dimension.

\subsection{Clone Detection}

After collecting the vectors of all functions, we have two applications to use them for code clone detection.
The first is to apply code clone detection by directly computing the similarity of these vectors.
More importantly, these vectors can also be used to enhance the detection effectiveness of other vector-based code clone detectors.
Figure \ref{fig:app} describes an example of the two applications of \emph{Gitor}.


\subsubsection{Application 1: Detect clones with global features.}

\par Cosine similarity is a commonly used metric, which measures similarity between two vectors, especially in high-dimension space. 
It measures similarity as the cosine of the angle between two vectors. 
Two similar vectors are expected to have a small angle between them. 
The cosine similarity of two vectors x and y is defined as follows:
\begin{equation*}
    \cos{\theta} = \frac{\Sigma_i^d x_i y_i}{\sqrt{\Sigma_i^d x_i^2} \sqrt{\Sigma_i^d y_i^2}}
\end{equation*}

Our first application is simply calculating the cosine similarity between two vectors,
If the similarity is greater than a certain threshold (\eg 0.7), they are identified as a clone pair, as illustrated in Figure \ref{fig:app}.






\subsubsection{Application 2: Detect clones with global and self features.}
\par We choose the graph as the representation of the whole code base since the natural structure of the graph can capture the underlying global connections among different code samples better than analyzing them individually. 
Instead of using the \emph{Gitor} alone, we can combine it with other self-features-based (\ie individual-features-based) methods, which is generated by individual analysis on each code sample. 
Nowadays, there are numerous vector-based code clone detection methods, such as \cite{wang2022heloc} and \cite{yokoi2018investigating}, and the current methods all focus on detecting clones utilizing individual features. 
In other words, our proposed \emph{global graph} based clone detection method can be combined with current vector-based detection methods to boost the performance of clone detection.

In this paper, we choose Doc2vec \cite{doc2vec}, Word2vec \cite{word2vec}, and Code2vec \cite{alon2019code2vec} as the vector-based detection methods \cite{wang2022heloc,yokoi2018investigating}. 
Word2vec \cite{word2vec} and Doc2vec \cite{doc2vec} are well-known natural language processing baseline methods for extracting feature vectors from source code. 
Code2vec \cite{alon2019code2vec} parses a code fragment into an AST path collection. 
To predict the method name, the core idea is to use a soft-attention mechanism on the paths and aggregate all vector representations into a single vector.
This combined method is illustrated in Figure \ref{fig:app}. The global vector and individual vector are added to calculate the similarity between two code samples.

%% file: outline/evaluation.tex
\section{EXPERIMENTS}

\par In this section, we aim to answer the following research questions:
    
\begin{itemize}
  \item \emph{RQ1: What is the effectiveness of Gitor in detecting different types of code clones when used alone?}
  \item \emph{RQ2: How does the use of global features contribute to the effectiveness of boosting individual-features-based clone detection?}
  \item \emph{RQ3: What is the effectiveness of Gitor compared to other state-of-the-art code clone detectors?}
  \item \emph{RQ4: What is the runtime performance of Gitor compared to other state-of-the-art clone detectors?}
\end{itemize}

\begin{table*}[htbp]
  \centering
  \small
  \caption{{Detection performance of \emph{Gitor} with different cosine similarity thresholds.}}
    \begin{tabular}{|c|cccc|cccc|cccc|}
    \toprule
    \multirow{2}[4]{*}{\textbf{Cosine = 0.60}} & \multicolumn{4}{c|}{\textbf{Keywords}} & \multicolumn{4}{c|}{\textbf{Side Information}} & \multicolumn{4}{c|}{\textbf{Both}} \\
\cmidrule{2-13}          & \textbf{16} & \textbf{32} & \textbf{64} & \textbf{128} & \textbf{16} & \textbf{32} & \textbf{64} & \textbf{128} & \textbf{16} & \textbf{32} & \textbf{64} & \textbf{128} \\
    \midrule
    \textbf{T-1 Recall} & 1     & 1     & 1     & 1     & 1     & 1     & 1     & 1     & 1     & 1     & 1     & 1 \\
    \textbf{T-2 Recall} & 1     & 1     & 1     & 1     & 1     & 1     & 1     & 1     & 1     & 1     & 1     & 1 \\
    \textbf{VST-3 Recall} & \multicolumn{1}{l}{0.986} & \multicolumn{1}{l}{0.972} & \multicolumn{1}{l}{0.991} & \multicolumn{1}{l|}{0.990} & 1 & 1 & 1 & \multicolumn{1}{l|}{0.999} & \multicolumn{1}{l}{0.988} & \multicolumn{1}{l}{0.989} & \multicolumn{1}{l}{0.993} & \multicolumn{1}{l|}{0.992} \\
    \textbf{ST-3 Recall} & \multicolumn{1}{l}{0.858} & \multicolumn{1}{l}{0.822} & \multicolumn{1}{l}{0.868} & \multicolumn{1}{l|}{0.847} & \multicolumn{1}{l}{0.995} & \multicolumn{1}{l}{0.995} & \multicolumn{1}{l}{0.995} & \multicolumn{1}{l|}{0.994} & \multicolumn{1}{l}{0.843} & \multicolumn{1}{l}{0.829} & \multicolumn{1}{l}{0.922} & \multicolumn{1}{l|}{0.906} \\
    \textbf{MT-3 Recall} & \multicolumn{1}{l}{0.638} & \multicolumn{1}{l}{0.617} & \multicolumn{1}{l}{0.560} & \multicolumn{1}{l|}{0.568} & \multicolumn{1}{l}{0.959} & \multicolumn{1}{l}{0.958} & \multicolumn{1}{l}{0.956} & \multicolumn{1}{l|}{0.952} & \multicolumn{1}{l}{0.672} & \multicolumn{1}{l}{0.683} & \multicolumn{1}{l}{0.780} & \multicolumn{1}{l|}{0.690} \\
    \textbf{Type-4 Recall} & \multicolumn{1}{l}{0.168} & \multicolumn{1}{l}{0.186} & \multicolumn{1}{l}{0.099} & \multicolumn{1}{l|}{0.089} & \multicolumn{1}{l}{0.606} & \multicolumn{1}{l}{0.595} & \multicolumn{1}{l}{0.559} & \multicolumn{1}{l|}{0.550} & \multicolumn{1}{l}{0.162} & \multicolumn{1}{l}{0.199} & \multicolumn{1}{l}{0.204} & \multicolumn{1}{l|}{0.114} \\
    \textbf{Precision} & \multicolumn{1}{l}{0.858} & \multicolumn{1}{l}{0.874} & \multicolumn{1}{l}{0.926} & \multicolumn{1}{l|}{0.919} & \multicolumn{1}{l}{0.687} & \multicolumn{1}{l}{0.694} & \multicolumn{1}{l}{0.712} & \multicolumn{1}{l|}{0.711} & \multicolumn{1}{l}{0.867} & \multicolumn{1}{l}{0.903} & \multicolumn{1}{l}{0.912} & \multicolumn{1}{l|}{0.936} \\
    \textbf{F1} & \multicolumn{1}{l}{0.651} & \multicolumn{1}{l}{0.655} & \multicolumn{1}{l}{0.625} & \multicolumn{1}{l|}{0.621} & \multicolumn{1}{l}{0.747} & \multicolumn{1}{l}{0.749} & \multicolumn{1}{l}{0.752} & \multicolumn{1}{l|}{0.750} & \multicolumn{1}{l}{0.659} & \multicolumn{1}{l}{0.684} & \multicolumn{1}{l}{0.715} & \multicolumn{1}{l|}{0.669} \\
    \midrule
    \textbf{Cosine = 0.70} & \textbf{16} & \textbf{32} & \textbf{64} & \textbf{128} & \textbf{16} & \textbf{32} & \textbf{64} & \textbf{128} & \textbf{16} & \textbf{32} & \textbf{64} & \textbf{128} \\
    \midrule
    \textbf{T-1 Recall} & 1     & 1     & 1     & 1     & 1     & 1     & 1     & 1     & 1     & 1     & 1 & 1 \\
    \textbf{T-2 Recall} & 1     & 1     & 1     & 1     & 1     & 1     & 1     & 1     & 1     & 1     & 1 & 1 \\
    \textbf{VST-3 Recall} & 0.979 & 0.930 & 0.958 & 0.946 & 1 & 1 & 1 & 0.999 & 0.964 & 0.954 & 0.988 & 0.989 \\
    \textbf{ST-3 Recall} & 0.811 & 0.753 & 0.777 & 0.761 & 0.995 & 0.995 & 0.994 & 0.993 & 0.755 & 0.750 & 0.840 & 0.803 \\
    \textbf{MT-3 Recall} & 0.537 & 0.486 & 0.401 & 0.402 & 0.945 & 0.950 & 0.945 & 0.943 & 0.548 & 0.537 & 0.601 & 0.492 \\
    \textbf{Type-4 Recall} & 0.101 & 0.099 & 0.045 & 0.039 & 0.498 & 0.486 & 0.446 & 0.436 & 0.097 & 0.103 & 0.090 & 0.046 \\
    \textbf{Precision} & 0.905 & 0.922 & 0.955 & 0.958 & 0.728 & 0.730 & 0.738 & 0.739 & 0.917 & 0.939 & 0.951 & 0.964 \\
    \textbf{F1} & 0.612 & 0.597 & 0.557 & 0.554 & 0.748 & 0.747 & 0.742 & 0.740 & 0.613 & 0.616 & 0.637 & 0.589 \\
    \midrule
    \textbf{Cosine = 0.80} & \textbf{16} & \textbf{32} & \textbf{64} & \textbf{128} & \textbf{16} & \textbf{32} & \textbf{64} & \textbf{128} & \textbf{16} & \textbf{32} & \textbf{64} & \textbf{128} \\
    \midrule
    \textbf{T-1 Recall} & 1     & 1     & 1     & 1     & 1     & 1     & 1     & 1     & 1     & 1     & 1     & 1 \\
    \textbf{T-2 Recall} & 1     & 1     & 1     & 1     & 1     & 1     & 1     & 1     & 1     & 1     & 1     & 1 \\
    \textbf{VST-3 Recall} & 0.936 & 0.871 & 0.918 & 0.907 & 1 & 0.999 & 0.999 & 0.999 & 0.897 & 0.939 & 0.963 & 0.931 \\
    \textbf{ST-3 Recall} & 0.721 & 0.655 & 0.620 & 0.634 & 0.995 & 0.994 & 0.994 & 0.993 & 0.681 & 0.672 & 0.700 & 0.666 \\
    \textbf{MT-3 Recall} & 0.425 & 0.312 & 0.233 & 0.231 & 0.916 & 0.911 & 0.903 & 0.898 & 0.422 & 0.370 & 0.378 & 0.290 \\
    \textbf{Type-4 Recall} & 0.051 & 0.040 & 0.015 & 0.013 & 0.343 & 0.329 & 0.296 & 0.287 & 0.048 & 0.040 & 0.027 & 0.014 \\
    \textbf{Precision} & 0.954 & 0.960 & 0.981 & 0.982 & 0.769 & 0.766 & 0.770 & 0.769 & 0.954 & 0.973 & 0.979 & 0.985 \\
    \textbf{F1} & 0.563 & 0.518 & 0.480 & 0.480 & 0.729 & 0.723 & 0.716 & 0.712 & 0.558 & 0.541 & 0.541 & 0.504 \\
    \bottomrule
    \end{tabular}%
  \label{tab:alone-results}%
\end{table*}%

\subsection{Experimental Settings}
\subsubsection{Dataset.}
We conduct our evaluations on the dataset: BigCloneBench \cite{big}, which consists of more than 8,000,000 labeled clone pairs from 25,000 systems. 
The code granularity of clone pairs in BigCloneBench \cite{big} is function-level, and each clone pair is manually assigned a corresponding clone type. 
Type-3 and Type-4 types are usually further divided into four subcategories based on their syntactical similarity score, as follows: 
i) \emph{Very Strongly Type-3} (VST3) with a similarity between [0.9, 1.0),
ii) \emph{Strongly Type-3} (ST3) with a similarity between [0.7, 0.9), 
iii) \emph{Moderately Type-3} (MT3) with a similarity between [0.5, 0.7), 
and iv) \emph{Weakly Type-3/Type-4} (WT3/T4) with a similarity between [0.0, 0.5).
The total number of these clone pairs used in our experiments is 8,446,574 including 8,139 Type-1 clones, 3,292 Type-2 clones, 4,577 VST3 clones, 3,469 ST3 clones, 7,606 MT3 clones, and 8,424,068 WT3/T4 clones. 
In the following experiment results, we use Type-4 (T4) to denote WT3/T4.

\subsubsection{Implementation.}

For \emph{individual information} extraction, we leverage a java parser (\ie \emph{javalang} \cite{javalang}) to extract the keywords and \emph{side information} from source code samples.
For global sample graph construction, we use a Python library, \emph{networkx} \cite{networkx}, to build a weighted and directed graph. 
For node embedding, we employ a widely used embedding method, \emph{ProNE} \cite{zhang2019prone}, to conduct the embedding of the graph. 
The output of embedding is a series of vectors of all nodes in the graph. 

We also select certain state-of-the-art code clone detection tools as our comparative systems, including \emph{SourcererCC}~\cite{sajnani2016sourcerercc}, \emph{CCFinder}~\cite{kamiya2002ccfinder}, \emph{NiCad}~\cite{roy2008nicad}, \emph{Deckard}~\cite{jiang2007deckard}, \emph{CCAligner}~\cite{wang2018ccaligner}, \emph{Oreo} \cite{saini2018oreo}, \emph{LVMapper} \cite{wu2020lvmapper}, and \emph{NIL} \cite{nakagawa2021nil}. 
All experiments are conducted on a server with Intel Xeon E5-2678 v3 @ 2.50GHz, 32 Gig-bytes memory, GeForce RTX 2080 TI Graphics Card and Ubuntu 18.04.5 LTS. 

\subsubsection{Metrics.}
We make use of the following widely used metrics to measure the detection performance of \emph{Gitor}.
Precision is defined as $P = TP/(TP+FP)$.
Recall is defined as $R = TP/(TP+FN)$.
F1 is defined as $F1 = 2*P*R/(P+R)$.
Among them, \emph{true positive} (TP) represents the number of samples correctly classified as clone pairs, \emph{false positive} (FP) represents the number of samples incorrectly classified as clone pairs, and \emph{false negative} (FN) represents the number of samples incorrectly classified as non-clone pairs.

\subsection{RQ1: Effectiveness of Gitor Used Alone \label{5.3}}

To examine the capability of \emph{Gitor} on clode clone detection, we conduct experiments from two perspectives, one is testing the performance of \emph{Gitor} alone, and another is testing the performance of \emph{Gitor} combined with other individual-features-based methods. 
In this part, we focus on checking the ability of \emph{Gitor} alone.
Specifically, we select different cosine similarity thresholds (\ie 0.6, 0.7, and 0.8) and different dimensions (\ie 16, 32, 64, and 128) of node embedding vectors to commence our evaluations. 
The results are illustrated in Table~\ref{tab:alone-results}, including the recall, precision, and F1 scores of our experiment on BigCloneBench dataset.
As for the measurement of precision, similar to other previous works \cite{wang2018ccaligner, saini2018oreo}, we randomly sample 400 clone pairs from clone reports in each tool and conduct manual analysis to validate them.
Each clone pair is checked independently by two experts. 
If there is a conflict, a final decision will be made after discussion with another expert. 
The principle rule for judging is based on the overall similarity between the two clone fragments and on whether they perform similar functionality. 

Through the results in Table \ref{tab:alone-results}, we find several interesting phenomena.
First, when the similarity threshold is different, the detection performance of \emph{Gitor} is also different. 
Basically, the larger the threshold, the higher the precision, but the lower the recall.
It is reasonable because the larger the threshold, the higher the similarity of the detected clones, and the higher the similarity is, the greater the probability of clones. 
But at the same time, some pairs whose similarity is slightly lower than the threshold will be filtered out, resulting in lower recall.
Second, when the vector dimensions are different, the detection performance of \emph{Gitor} is also different. 
This is normal because the dimensions of the vectors are different, the degree of retention of graph information will also be different. 
Basically, when we combine \emph{keywords} with \emph{side information}, the larger the dimension of the vector, the higher the precision.
Third, the features obtained when selecting \emph{keywords} to construct a sample graph are more accurate than when selecting \emph{side information}.
In other words, when using \emph{keywords} to construct the global graph, the precision of \emph{Gitor} is higher than when selecting \emph{side information}. 
This is because \emph{keywords} represent the key tokens in the programming language, these key tokens are not allowed to be changed, and different key tokens describe different program information. 
\emph{Gitor} can preserve more program semantics when all keywords are considered.
Forth, after combining \emph{keywords} with \emph{side information}, \emph{Gitor}'s precision is mostly improved. 
It shows that the combination of the two information allows \emph{Gitor} to retain more program semantics. 
At this time, when the vector dimension is 64, the average F1 under the three thresholds (\ie 0.6, 0.7, and 0.8) is the highest.

Based on the above findings, we suggest that if researchers want to detect more clones, they can set the threshold to 0.6 when using \emph{Gitor} with \emph{keywords} and \emph{side information}.
In addition, if researcher like to detect clones with higher accuracy, they can set the threshold to a higher value, such as 0.7 or 0.8.

\begin{table}[htbp]
  \centering
  \small
  \caption{{Detection performance of individual-features-based detectors}}
\begin{tabular}{c|c|c|c}
    \toprule
    \toprule
    \multicolumn{1}{c}{\multirow{2}[4]{*}{Method}} & \multicolumn{3}{c}{ Individual-features-based detector} \\
\cmidrule{2-4}    \multicolumn{1}{c}{} & Doc2Vec & W2V-avg & Code2Vec \\
    \midrule
    Type-1 Recall & 1 & 1 & 1 \\
    Type-2 Recall & 0.95 & 1 & 1 \\
    VST-3 Recall & 0.86 & 0.99 & 0.998 \\
    ST-3 Recall & 0.57 & 0.85 & 0.995 \\
    MT-3 Recall & 0.21 & 0.53 & 0.979 \\
    Type-4 Recall & 0.02 & 0.11 & 0.928 \\
    Precision & 0.98 & 0.98 & 0.619 \\
    F1 & 0.47 & 0.63 & 0.753 \\
    \bottomrule
    \bottomrule
    \end{tabular}%
  \label{tab:ii1}%
\end{table}%

\begin{table}[htbp]
  \centering
  \small
  \caption{{Detection performance of individual-features-based detectors combined with \emph{Gitor}}}
    \begin{tabular}{c|c|c|c}
    \toprule
    \toprule
    \multicolumn{1}{c}{\multirow{2}[4]{*}{Method}} & \multicolumn{3}{c}{With Gitor (Ours)} \\
\cmidrule{2-4}    \multicolumn{1}{c}{} & Doc2Vec & W2V-avg & Code2Vec \\
    \midrule
    Type-1 Recall & 1 & 1 & 1 \\
    Type-2 Recall & 1 & 1 & 1 \\
    VST-3 Recall & 0.947 & 0.973 & 0.998 \\
    ST-3 Recall & 0.671 & 0.817 & 0.996 \\
    MT-3 Recall & 0.421 & 0.573 & 0.979 \\
    Type-4 Recall & 0.042 & 0.131 & 0.939 \\
    Precision & 0.971 & 0.981 & 0.65 \\
    F1 & 0.558 & 0.649 & 0.778 \\
    \bottomrule
    \bottomrule
    \end{tabular}%
  \label{tab:cm1}%
\end{table}%

\subsection{RQ2: Combination with Other Individual Features-based Methods\label{rq2exp}}
\par In this part, we pay attention to the effectiveness when \emph{Gitor} is combined with other detection methods. 
Since our system is purely based on global features, we first want to explore how would it contribute to current individual-features-based detection methods. 
In order to check the effectiveness of detection using the combination of global features and individual features, we pick several widely used methods \cite{wang2022heloc,yokoi2018investigating}, which have been proved effective on clone detection, then we test the effectiveness when they are combined with \emph{Gitor}. 

\par In this experiment, we choose Doc2Vec \cite{doc2vec}, Word2Vec \cite{word2vec}, and Code2Vec \cite{alon2019code2vec} as the individual-features-based detection methods \cite{wang2022heloc,yokoi2018investigating}. 
We define the average vectors of Word2Vec as W2V-avg, and the Doc2Vec extends the word vectors to entire document vectors. 
The Code2Vec can embed the entire code sample into a single vector. We choose \emph{cosine similarity} as the similarity metric in this part of experiments, and we use the default parameters for Doc2Vec, Word2Vec, and Code2Vec.

\begin{table*}[htbp]
  \centering
  \small
  \caption{{Detection performance of \emph{SourcererCC}~\cite{sajnani2016sourcerercc}, \emph{CCFinder}~\cite{kamiya2002ccfinder}, \emph{NiCad}~\cite{roy2008nicad}, \emph{Deckard}~\cite{jiang2007deckard}, \emph{CCAligner}~\cite{wang2018ccaligner}, \emph{Oreo}~\cite{saini2018oreo}, \emph{LVMapper}\cite{wu2020lvmapper}, \emph{NIL}\cite{nakagawa2021nil}, and \emph{Gitor} on detecting different types of code clones.}}
    \begin{tabular}{cccccccccc}
    \toprule
    \toprule
    \textbf{Tool} & SourcererCC & CCFinder & NiCad & Deckard & CCAligner & Oreo & LVMapper & NIL & Gitor \\
    \midrule
    \textbf{Type-1 Recall} & 1 & 1 & 1 & 0.6 & 1 & 1 & 0.99 & 0.99 & \textbf{1} \\
    \textbf{Type-2 Recall} & 0.97 & 0.93 & 0.99 & 0.58 & 0.99 & 0.99 & 0.99 & 0.96 & \textbf{1} \\
    \textbf{Very Strongly Type-3 Recall} & 0.93 & 0.62 & 0.98 & 0.62 & 0.97 & 1 & 0.98 & 0.93 & \textbf{0.988} \\
    \textbf{Strongly Type-3 Recall} & 0.6 & 0.15 & \textbf{0.93} & 0.31 & 0.7 & 0.89 & 0.81 & 0.67 & 0.84 \\
    \textbf{Moderately Type-3 Recall} & 0.05 & 0.01 & 0.008 & 0.12 & 0.1 & 0.3 & 0.19 & 0.1 & \textbf{0.601} \\
    \textbf{Type-4 Recall} & 0 & 0 & 0 & 0.01 & - & 0.007 & - & - & \textbf{0.09} \\
    \midrule
    \textbf{Precision} & 0.978 & 0.72 & \textbf{0.99} & 0.348 & 0.8 & 0.895 & 0.58 & 0.94 & 0.951 \\
    \bottomrule
    \bottomrule
    \end{tabular}%
  \label{tab:gcj1}%
\end{table*}%

\par To evaluate the detection performance of these three methods, we test them on the BigCloneBench dataset. 
We use three methods to get the embeddings of all code samples and compare the \emph{cosine similarity} to detect possible clone pairs, where we set the similarity threshold as 0.9, and embedding dimension as 128 since these tools reach their best performance under this setting \cite{yokoi2018investigating}. 
Table~\ref{tab:ii1} shows the detection results including recall, precision and F1-score on the BigCloneBench dataset. 
Then we combine the vectors generated by the above methods with \emph{Gitor} generated vectors and conduct the similarity comparison on the BigCodeBench dataset, where the similarity threshold is set to 0.9 as well and the dimension of \emph{Gitor} is 32 since \emph{Gitor} performs the best with dimension as 32 when the similarity threshold set to 0.9. 
The results are illustrated in Table~\ref{tab:cm1}, where we can see that the overall detection performance, including recall, precision, and F1-score, is significantly improved compared to the original results, so it suggests that \emph{Gitor} can boost the effectiveness of other individual-features-based detection methods.

\par In short, the \emph{Gitor} is not only effective when used alone, but also able to boost the performance of other individual-features-based detection methods.

\subsection{RQ3: Comparative with Other Detectors}

In order to evaluate \emph{Gitor}'s performance comprehensively, we compare the performance of \emph{Gitor}'s clone detection against the latest versions of several publicly available clone detection tools, such as \emph{SourcererCC}~\cite{sajnani2016sourcerercc}, \emph{CCFinder}~\cite{kamiya2002ccfinder}, \emph{NiCad}~\cite{roy2008nicad}, \emph{Deckard}~\cite{jiang2007deckard}, \emph{CCAligner}~\cite{wang2018ccaligner}, \emph{Oreo} \cite{saini2018oreo}, \emph{LVMapper} \cite{wu2020lvmapper}, and \emph{NIL} \cite{nakagawa2021nil}. 
Since most of traditional code clone detection tools (\eg \emph{SourcererCC}~\cite{sajnani2016sourcerercc} and \emph{NiCad}~\cite{roy2008nicad}) select 0.7 as their thresholds to identify code clones, we also choose 0.7 as the threshold to commence our comparative evaluations.
Through the results in Table \ref{tab:alone-results}, we observe that \emph{Gitor} can maintain the best overall performance (\ie F1) when the dimension of node embedding vectors is 64.
Therefore, we use the corresponding detection results as the comparative performance of \emph{Gitor},
\emph{SourcererCC}, \emph{CCFinder}, \emph{Nicad} , \emph{Deckard}, \emph{CCAligner}, \emph{Oreo}, \emph{LVMapper}, and \emph{NIL}, where the recall numbers are summarized per clone category. 
As Table \ref{tab:gcj1} shows, \emph{Gitor} outperforms every other tool on most of the clone categories, except for ST3. 
Although NiCad performs the best on ST3, \emph{Gitor}'s performance on ST3 clone is still quite comparable to the state-of-art since there is only a 9 percent difference. 
The recall results are promising since they suggest that, in addition to recognizing easier-to-find clones like T1, T2, and VST3, \emph{Gitor} also detects clones that other tools miss. 
In comparison to other methods, where Oreo's highest recall is 0.3, 0.601 recall in the MT3 category is a significant improvement. 
Table~\ref{tab:gcj1} also shows the precision results of all tools. The precision of \emph{Gitor} is 0.951, and only SourcererCC and NiCard perform marginally better than \emph{Gitor} (by 5 percent).

The recall and precision experiments show that \emph{Gitor} is a reliable and accurate clone detector that can detect Type-1, Type-2, and Type-3 clones efficiently and detect part of Type-4 clones. To address this issue, in the future we will improve \emph{Gitor} with better chosen individual information to detect Type-4 clones more effectively.

\begin{table*}[htbp]
  \centering
  \small
  \caption{{Runtime performance of \emph{SourcererCC}~\cite{sajnani2016sourcerercc}, \emph{CCFinder}~\cite{kamiya2002ccfinder}, \emph{NiCad}~\cite{roy2008nicad}, \emph{Deckard}~\cite{jiang2007deckard}, \emph{CCAligner}~\cite{wang2018ccaligner}, \emph{Oreo}~\cite{saini2018oreo}, \emph{LVMapper}\cite{wu2020lvmapper}, \emph{NIL}\cite{nakagawa2021nil}, and \emph{Gitor}.}}
    \begin{tabular}{cccccccccc}
    \toprule
    \toprule
    LOC & SourcererCC & CCFinder & NiCad & Deckard & CCAligner & Oreo & LVMapper & NIL & Gitor \\
    \midrule
    \textbf{1K} & 3s & 2s & 1s & 1s & 1s & 1s & 1s & 1s & \textbf{0.03s} \\
    \textbf{10K} & 5s & 5s & 2s & 4s & 2s & 3s & - & - & \textbf{0.18s} \\
    \textbf{100K} & 7s & 10s & 5s & 32s & 3s & 6s & - & - & \textbf{1.26s} \\
    \textbf{1M} & 37s & 39s & 12s & 27m12s & 11m52s & 4m34s & 29s & \textbf{10s} & 13.10s \\
    \textbf{10M} & 12m21s & 6m30s & 19m49s & Killed & 29m48s & 36m6s & 13m 38s & \textbf{1m 38s} & 2m11s \\
    \textbf{100M} & 12h27m & 9h49m & Killed & - & Killed & 1d13h46m & 17h 23m 39s & 1h 38m 29s & \textbf{1h7min} \\
    \midrule
    \midrule
    \multicolumn{10}{c}{Killed means the tool fails to parse the code or report out-of-memory errors, "-" means no such data in previous study} \\
    \end{tabular}%
  \label{tab:rt}%
\end{table*}%

\subsection{RQ4: Scalability}
In this section, we pay attention on the runtime performance of \emph{Gitor}. As mentioned before, scalability is an important requirement for clone detection methods, and \emph{Gitor} is designed as a scalable clone detection system. So, we will evaluate the efficiency and demonstrate the scalability of \emph{Gitor} in two parts: training efficiency and classification efficiency.

\textbf{Dataset for scalability experiments:} We use the whole dataset of BigCloneBench (\ie IJaDataset \cite{ijdata}), which is a widely used dataset containing about 250 million lines of Java source code mined from SourceForge and Google Code. 
The full IJaDataset and its subsets are often used for evaluating execution
time and scalability of clone detection tools \cite{sajnani2016sourcerercc, li2020saga, wang2018ccaligner}. 
We test \emph{Gitor} using inputs with different sizes generated from this dataset.

\textbf{Different sizes for scalability experiments:} Execution time primarily depends on the size of the input in terms of the number of lines of code (LOC) needed to be processed and classified by the system. 
So, we build the inputs with varying convenient sizes (\ie 1K, 10K, 100K, 1M, 10M, and 100M LOC) by randomly selecting samples from IJaDataset.

\textbf{Results:} The execution is finished on a machine with Intel Xeon E5-2678 v3 @ 2.50GHz 12 cores CPU, 32GB of memory, GeForce RTX 2080 TI Graphics Card, and system is Ubuntu 18.04.5 LTS. 
For \emph{Gitor}, it mainly consists of two phases, the first it to apply node embedding to extract all functions' vectors, and then these vectors will be used to compute cosine similarity one by one.
In practice, it takes little time to complete the first phase (\ie 20 minutes for 100M LOC).
However, when the code size becomes large, the number of functions will also be large, resulting in a massive number of code pairs to be analyzed.
To mitigate the issue, we adopt matrix computation to calculate the similarity of all code pairs, where GPU is used to accelerate the computation process. The runtime of \emph{Gitor} is the total time of two phases included.
The runtime performance of all the above tools and corresponding LOC are listed in Table~\ref{tab:rt}, which shows that \emph{Gitor} outperforms the seven state-of-the-art clone detection tools in all sizes of inputs while \emph{Gitor} is a bit slower than \emph{NIL} in 1MLOC and 10MLOC size, but \emph{Gitor} is still more efficient than the state-of-the-art detector \emph{NIL} when it comes to larger size, 100MLOC, in this case.

In conclusion, \emph{Gitor} is eight times faster than the token-based detection tool \emph{CCFinder}~\cite{kamiya2002ccfinder} with the input size of 100 million LOC, which means it is highly scalable.

\subsection{Summarization}
Our experimental results demonstrate \emph{Gitor}'s effectiveness as a code clone detection method.
It achieves optimal accuracy using a combination of keyword and side information features (RQ1). 
\emph{Gitor} improves the performance of individual feature-based detectors when used jointly (RQ2). 
In comparative evaluations, \emph{Gitor} attains higher recall than eight state-of-the-art tools on the BigCloneBench dataset, with precision comparable to top techniques (RQ3).
Moreover, \emph{Gitor} analyzes 100 million lines of code efficiently in just 1 hour, and is the fastest tool on large code bases, running 100X faster than CCFinder (RQ4). 
In summary, through extensive evaluations, our results consistently highlight \emph{Gitor}'s strengths in terms of effectiveness, enhancement capability, superior accuracy over current methods, and scalability to large code bases.

%% file: outline/discussion.tex
\section{DISCUSSION}

\par \noindent \emph{\textbf{{Why Gitor outperforms the other approaches}}}. 
First, currently existing clone detection tools (\eg \emph{CCFinder} \cite{kamiya2002ccfinder} and \emph{SourcererCC} \cite{sajnani2016sourcerercc}) focus on analyzing code samples individually without considering the underlying connection among code samples. 
However, \emph{Gitor} considers the connection among different code samples by extracting the \emph{individual information} of a code sample as its representation, and the extracted \emph{individual information} is used to build a global graph to represent the whole code base which preserves the underlying connections of all code samples.
\par \noindent \emph{\textbf{{Why not compare with deep learning based methods}}}. 
First of all, \emph{Gitor} is not a deep learning-based method, and the experiment we conduct in Section \ref{rq2exp} is only used to prove that \emph{Gitor} can boost the performance of other detection methods, which does not suggest that \emph{Gitor} is a deep learning based. 
Second, deep learning-based methods require training a detector on large labeled datasets, which is time-consuming and limits the practicability and scalability of deep learning-based clone detectors. In contrast, \emph{Gitor} does not require time-consuming training on large labeled datasets, making it more practical for use in real-world applications. 


\par \noindent \emph{\textbf{{Future work}}}. The embedding process of \emph{Gitor} is very efficient since we make use of \emph{ProNE}~\cite{zhang2019prone}, however, the code clone classification process is quite time-consuming due to its $\mathcal{O}(n^2)$ complexity to get all clones detected. In future work, we consider techniques like filtering to improve our classification speed by filtering most of the unlikely code pairs according to the properties of the sample itself, such as lines of a sample, and the number of tokens in a sample. Besides, to achieve better detection performance, we will explore more types of \emph{individual information} to represent code samples more properly and more accurately.

%% file: outline/relatedwork.tex
\section{RELATED WORK}

\par This section introduces related studies on code clone detection, which can be classified into five categories: text-based methods, token-based methods, tree-based methods, graph-based methods, and metrics-based methods.

\par The similarity between two code snippets is measured in the form of text or strings for the text-based methods \cite{johnson1994substring, ducasse1999language, roy2008nicad}. \cite{johnson1994substring} proposes a fingerprinting technique for detecting code clones. \cite{ducasse1999language} develops a language-independent method for detecting similar codes using only line-based string matching. These two techniques, however, do not support Type-3 clone detection. To detect more types of clones, \emph{Nicad} \cite{roy2008nicad} introduces a two-stage approach that consists of i) identifying and normalizing potential clones using flexible pretty-printing and ii) computing similarity by simply text-line comparison using the longest common subsequence algorithm. Although \emph{Nicad} can detect a number of Type-3 clones, it cannot detect Type-4 clones because it ignores the program semantics of given code samples.

\par For the token-based techniques \cite{kamiya2002ccfinder, gode2009incremental, sajnani2016sourcerercc, li2017cclearner, wang2018ccaligner}, tokens are first collected from program code by lexical analysis. \emph{CCFinder} \cite{kamiya2002ccfinder} extracts a token sequence from the input code and converts it into a regular form for finding Type-1 and Type-2 clones using numerous rule-based transformations, and \emph{SourcererCC} \cite{sajnani2016sourcerercc} has been developed to support Type-3 clone detection, which is designed to capture the tokens' overlap similarity among multiple approaches for detecting Type-3 clones that are close to being detected. \emph{SourcererCC} \cite{sajnani2016sourcerercc} is the most scalable code clone detector, capable of detecting 250 million lines of code. However, token-based detection methods, like text-based approaches, are unable to handle Type-4 clones.

\par To detect code clones, the tree-based tools \cite{jiang2007deckard, wei2017cdlh, zhang2019astnn} employ \emph{Abstract Syntax Tree} (AST) as the code representation. \emph{Deckard} \cite{jiang2007deckard}'s core idea is to compute characteristic vectors within ASTs and use \emph{Locality Sensitive Hashing} (LSH) to cluster comparable vectors for clone detection. \emph{CDLH} \cite{wei2017cdlh} first converts ASTs to binary trees, then uses Tree-LSTM \cite{tai2015treelstm} to encode these trees into vector representations.  Finally, these vectors are utilized to compare distinct codes' similarity. \emph{ASTNN} \cite{zhang2019astnn} separates each huge AST into a sequence of little statement trees, unlike \emph{CDLH} \cite{wei2017cdlh}. To find semantic code clones, after encoding these statement trees into vectors, a bidirectional RNN model is utilized to construct the final vector representation of a code fragment. These tree-based methods can detect semantic clones, but their scalability is limited due to their long execution times.

\par For the graph-based methods \cite{krinke2001duplix, komondoor2001pdgdup, wang2017ccsharp, zhao2018deepsim, chen2014centroid}, program semantics are first distilled into multiple graph representations, such as program dependency graph and control flow graph.
\cite{komondoor2001pdgdup} and \cite{krinke2001duplix} both extract program dependency graphs from code fragments and locate similar codes by excavating isomorphic subgraphs to represent code clones. \emph{CCSharp} \cite{wang2017ccsharp} employs two strategies to reduce the overall processing cost of \cite{komondoor2001pdgdup} and \cite{krinke2001duplix}: graph structure modification and characteristic vector filtering. However, due to the complexity of graph isomorphism and the heavy-weight time consumption of graph matching, it still has low scalability on large-scale code clone detection.

\par Metrics can be obtained from tree or graph representations of source code or straight from source code for the metrics-based techniques \cite{mayrand1996experiment, balazinska1999measuring, patenaude1999extending, saini2018oreo}. Both \cite{balazinska1999measuring} and \cite{mayrand1996experiment} use metrics extracted from the AST to describe the source code and to identify code clones. In addition, \cite{patenaude1999extending} detects clones using a variety of metrics collected from source code (\eg classes, coupling, and hierarchical organization). These approaches use code features to determine how similar two code fragments are in terms of semantics.

%% file: outline/conclusion.tex
\section{CONCLUSION}

\par In this paper, we propose \emph{Gitor} to achieve scalable code clone detection. 
Given a source code base, we first generate a global graph representing the whole code base, and then apply graph embedding to extract the vectors of all code samples in the code base.
Finally, the code sample vectors can be simply used to compute the similarity of different code sample. 
We evaluate \emph{Gitor} on a widely used dataset and compare \emph{Gitor} with other widely used code clone detection methods. 
The results show that \emph{Gitor} is superior to \emph{SourcererCC}~\cite{sajnani2016sourcerercc}, \emph{CCFinder}~\cite{kamiya2002ccfinder}, \emph{NiCad}~\cite{roy2008nicad}, \emph{Deckard}~\cite{jiang2007deckard}, \emph{CCAligner}~\cite{wang2018ccaligner}, and \emph{Oreo}~\cite{saini2018oreo} on both effectiveness and scalability. 
Moreover, \emph{Gitor} requires only about one hour to analyze 100 million lines of code and is the most scalable among our comparative tools.

\section{DATA AVAILABILITY}
Our data are available on our website: \href{https://github.com/Gitor-clone/Gitor}{https://github.com/Gitor-clone/Gitor}.

%% file: outline/acknow.tex
\section*{ACKNOWLEDGEMENTS}
We would thank the anonymous reviewers for their insightful comments to improve the quality of the paper.
This research/project is supported by the National Research Foundation, Singapore, and the Cyber Security Agency under its National Cybersecurity R\&D Programme (NCRP25-P04-TAICeN). 
Any opinions, findings and conclusions or recommendations expressed in this material are those of the author(s) and do not reflect the views of National Research Foundation, Singapore and Cyber Security Agency of Singapore.